\documentclass{article}
\usepackage{ksmi}
\usepackage{amsmath,cite,url}
\usepackage{graphicx}
\usepackage{color}
\usepackage{tikz}
\usetikzlibrary{arrows.meta,positioning,shapes.geometric,calc}
\usepackage{booktabs}

\title{PiAnnotate: A Web Annotation Tool for Piano Fingering, with a Diagnostic Probe}

\def\authorname{Joonhyung Bae, Kirak Kim, Hyeyoon Cho, Sein Lee, Yoon-Seok Choi, Hyeon Hur, Gyubin Lee, Akira Maezawa, Jonghwa Park, Jaebum Park, and Juhan Nam}

\makeatletter
\gdef\@author{%
  \begin{tabular}{@{}c@{}}
    \textbf{Joonhyung Bae$^{1}$, Kirak Kim$^{1}$, Hyeyoon Cho$^{1}$, Sein Lee$^{1}$}\\
    \textbf{Yoon-Seok Choi$^{2}$, Hyeon Hur$^{2}$, Gyubin Lee$^{1}$, Akira Maezawa$^{3}$}\\
    \textbf{Jonghwa Park$^{2}$, Jaebum Park$^{2}$, Juhan Nam$^{1,*}$}\\[0.4em]
    \normalsize $^{1}$Korea Advanced Institute of Science and Technology (KAIST), Daejeon, South Korea\\
    \normalsize $^{2}$Seoul National University, Seoul, South Korea\\
    \normalsize $^{3}$YAMAHA, Hamamatsu, Japan\\[0.2em]
    \small \texttt{\{jh.bae,kirak,hyeyooncho,seinlee,gbstorm81,juhan.nam\}@kaist.ac.kr}\\
    \small \texttt{\{dallas71,sunny000927,parkpe95\}@snu.ac.kr}\\
    \small \texttt{pianistpark@gmail.com}\\
    \small \texttt{akira.maezawa@music.yamaha.com}\\
    \normalsize $^{*}$Corresponding author
  \end{tabular}}
\makeatother

\begin{document}

\maketitle

%-----------------------------------------------------
\begin{abstract}
Piano fingering shapes how a passage can be played, yet it is difficult to label after a performance.
An annotator must decide which finger produced each note while reconciling the score, timing, video, and hand motion.
We present \textbf{PiAnnotate}, a web-based pipeline for adding expert fingering annotations to the F\"urElise performance dataset~\cite{wang2024furelise}.
The tool brings together a piano-roll view, performance video, and a 3D MANO hand mesh so that reviewers can inspect each assignment in musical and physical context.
Rather than storing only the final answer, PiAnnotate keeps paired rule-based and human-edited fingering tracks.
These paired tracks make the annotation history auditable by showing where a geometric rule was sufficient, where experts intervened, and how labels changed across review passes.
As a final diagnostic, we train a small Transformer probe on the paired tracks.
The probe improves on the rule baseline on held-out pieces while remaining conservative about changing labels that were already correct, suggesting that the edited labels contain learnable structure rather than only isolated fixes.
\end{abstract}

%-----------------------------------------------------
\section{Introduction}\label{sec:intro}
%-----------------------------------------------------

Many methods have tried to predict piano fingering, from ergonomic and physical rules~\cite{parncutt1997model,alkasimi2007physical} to HMMs, statistical learning, and neural models~\cite{yonebayashi2007automatic,nakamura2020statistical,ramoneda2022automatic}.
These methods all depend on trustworthy labels for which finger played each note, a simple resource that is expensive to create.
Existing fingering corpora such as PIG~\cite{nakamura2020statistical} are valuable, but they cover only part of the annotation need.
Recent multimodal datasets and toolkits capture richer performance evidence, including video and MIDI workflows~\cite{kim2025pianovam,park2025webtoolkits}.
F\"urElise~\cite{wang2024furelise} goes further by providing synchronized audio, video, MIDI, and 3D hand motion~\cite{simon2017hand,goebl2013temporal}, but it still does not say which finger played each note.

Turning motion data into fingering labels is still an expert annotation problem.
Reviewers need to compare note events with hand position and decide whether an assignment is physically plausible.
PiAnnotate is designed for this setting.
It does not replace expert review; it makes review easier to audit by placing motion evidence, rule labels, edited labels, and review-stage metadata on the same timeline.
Time-aligned tools exist in adjacent domains, including Praat~\cite{boersma2001praat}, ELAN~\cite{brugman2004elan}, and Sonic Visualiser~\cite{cannam2010sonic}, but they lack 3D hand-mesh rendering and finger-specific piano workflows.

This paper describes \textbf{PiAnnotate}, the pipeline we used to add expert fingering annotations to the full F\"urElise collection.
Our contributions are as follows.
\begin{enumerate}\itemsep0pt
  \item A web annotation tool with frame-level editing, 3D MANO hand mesh~\cite{romero2017embodied} visualization, and an explicit multi-stage review workflow (Sec.~\ref{sec:tool}).
  \item An expert-verified fingering layer for F\"urElise, organized as paired rule-based and human-edited tracks so that rule errors can be inspected directly (Sec.~\ref{sec:dataset}).
  \item A small Transformer-style diagnostic probe~\cite{huang2019music} that checks whether the edited labels contain learnable structure beyond isolated fixes and surfaces a timestamp-related annotation artifact (Sec.~\ref{sec:probe} -- \ref{sec:findings}).
\end{enumerate}
We borrow the term \emph{diagnostic probe} from NLP~\cite{belinkov2019probing}; to our knowledge, this is the first application to a music annotation corpus.
The model is not a deployable corrector because the full collection is used in training and the inference gate is a composite confidence filter.
The setup is based on human-in-the-loop annotation~\cite{settles2009active}, model-assisted relabeling~\cite{northcutt2021confident}, and learning from noisy labels~\cite{song2022learning}.

%-----------------------------------------------------
\section{Pipeline Overview}\label{sec:pipeline}
%-----------------------------------------------------

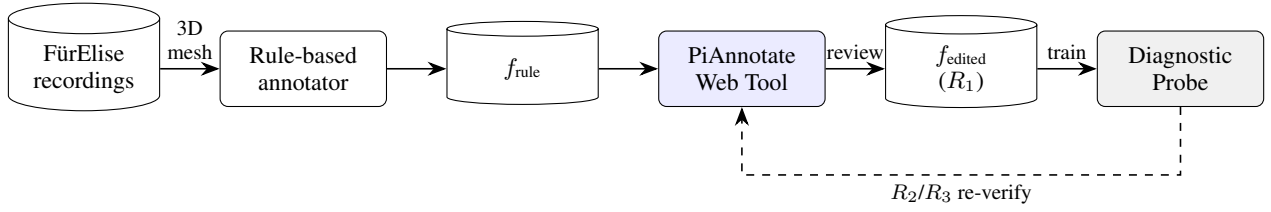
\begin{figure*}[t]
\centering
\begin{tikzpicture}[
    every node/.style={font=\small},
    box/.style={draw, rounded corners=3pt, align=center, inner sep=4pt, minimum height=10mm, minimum width=22mm},
    db/.style={draw, cylinder, shape border rotate=90, aspect=0.25, align=center, inner sep=4pt, minimum height=11mm, minimum width=20mm},
    arr/.style={-{Stealth[length=2.5mm]}, semithick},
    lbl/.style={font=\footnotesize, midway}
  ]
  \node[db]  (fur)   at (0,0)     {F\"urElise\\recordings};
  \node[box] (rule)  at (2.9,0)   {Rule-based\\annotator};
  \node[db]  (frule) at (5.8,0)   {$f_{\text{rule}}$};
  \node[box, fill=blue!8] (tool) at (8.7,0) {PiAnnotate\\Web Tool};
  \node[db]  (edit)  at (11.6,0)  {$f_{\text{edited}}$\\($R_1$)};
  \node[box, fill=black!6] (probe) at (14.5,0) {Diagnostic\\Probe};
  \draw[arr] (fur)   -- node[lbl, above, align=center]{3D\\mesh} (rule);
  \draw[arr] (rule)  -- (frule);
  \draw[arr] (frule) -- (tool);
  \draw[arr] (tool)  -- node[lbl, above]{review} (edit);
  \draw[arr] (edit)  -- node[lbl, above]{train} (probe);
  \draw[arr, dashed] (probe.south) -- ++(0,-0.9) -| node[lbl, below, pos=0.25]{$R_2$/$R_3$ re-verify} (tool.south);
\end{tikzpicture}
\caption{PiAnnotate pipeline. F\"urElise 3D hand-mesh recordings are processed by a geometric rule-based annotator to produce $f_{\text{rule}}$, which the annotator reviews and corrects in the web tool to produce $f_{\text{edited}}$. Edited pairs train a diagnostic probe whose output is optionally surfaced in subsequent review passes (dashed).}
\label{fig:pipeline}
\end{figure*}

\figref{fig:pipeline} sketches the loop.
A rule-based annotator extracts MIDI onset-offset intervals on the motion-frame grid.
For each pressed key at onset, it scans both hands' five MANO fingertips, keeps candidates inside the key's pitch and front-back bounds and near the key surface, and chooses the minimum score combining surface-height distance with normalized front-back distance.
If no fingertip passes the test, the rule label is missing.
The assigned label is propagated until the note offset to produce $f_{\text{rule}}$.
A human reviewer opens the piece in the PiAnnotate web tool, edits frames where the rule output is wrong, and marks the piece as $R_1$-checked in a per-piece status JSON.
Once a non-trivial $R_1$ pool exists, a probe Transformer is trained on $(f_{\text{rule}}, f_{\text{edited}})$ pairs and run on every piece; its output is written to a separate directory and optionally surfaced to the reviewer in subsequent passes ($R_2$, $R_3$).

%-----------------------------------------------------
\section{Annotation Tool}\label{sec:tool}
%-----------------------------------------------------

\begin{figure*}[t]
\centering
\includegraphics[width=0.95\textwidth]{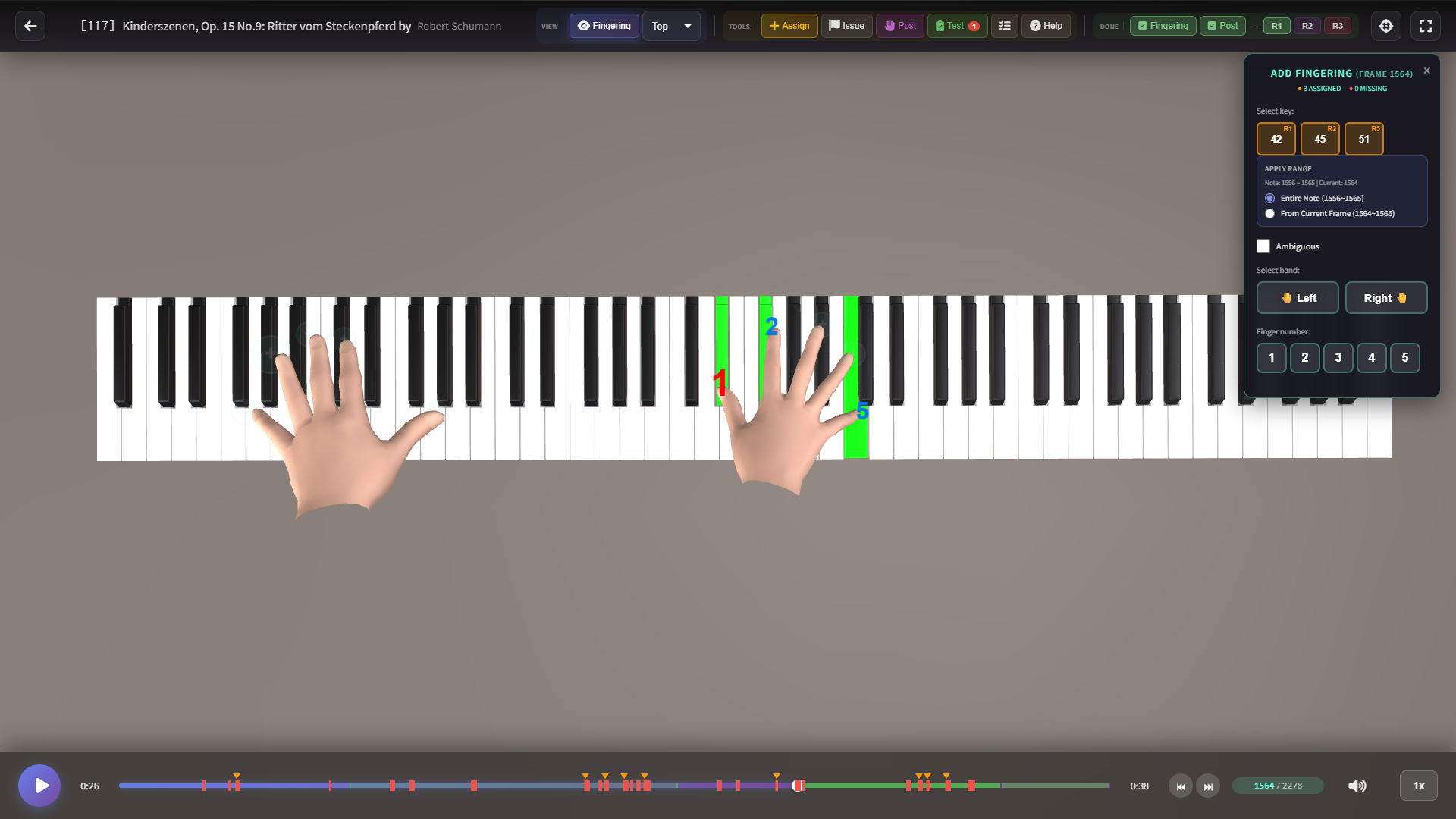}
\caption{The PiAnnotate web annotation tool. A top-down piano view renders the F\"urElise hand-mesh tracks in 3D for the current frame; pressed keys are highlighted in green and overlaid with their currently assigned finger numbers (red = left hand, blue = right). The right-hand panel adds a fingering at the playhead frame: the key, hand, and finger are selected with single clicks, and the assignment can be extended to the entire sustained note or applied only from the current frame onward. The bottom timeline shows the audio waveform with click-onset markers and the per-piece review state ($R_1$/$R_2$/$R_3$) is tracked in the top-right corner.}
\label{fig:tool}
\end{figure*}

The tool is a browser application: a Vite/React frontend talks to a Flask backend over a small REST API.
Each piece bundles the synced video, audio, MIDI, key-press timing, and per-frame 3D hand mesh provided by F\"urElise~\cite{wang2024furelise}; PiAnnotate adds the fingering layer on top (\figref{fig:tool}).
The annotator sees three coupled views: a piano roll with overlaid finger labels colored by hand; the original performance video; and a 3D MANO hand mesh rendered for the current frame, which makes physically implausible assignments (e.g., a thumb on a key the thumb cannot reach) immediately visible.
Editing is keyboard-driven (\tabref{tab:keys}): $\leftarrow/\rightarrow$ steps through fingering events, digit keys $1$--$5$ assign a finger, and \texttt{Space} toggles playback.

\begin{table}[h]
  \centering
  \small
  \begin{tabular}{ll}
    \toprule
    Key & Action \\
    \midrule
    \texttt{Space}       & Play / Pause \\
    $\leftarrow$ $\rightarrow$ & Previous / Next fingering event \\
    \texttt{1}--\texttt{5} & Assign finger number \\
    \texttt{F}           & Toggle fingering overlay \\
    \texttt{ESC}         & Deselect current fingering \\
    \bottomrule
  \end{tabular}
  \caption{Keyboard shortcuts in the annotation tool.}
  \label{tab:keys}
\end{table}
Edits are committed to an edited file per-piece.
Each piece carries a JSON status tracking three review stages ($R_1$, $R_2$, $R_3$) with timestamps; only $R_1$-checked pieces are admitted to probe training.
The tool also provides an in-browser backup/recovery path for unsaved edits, which has proven necessary in practice during multi-hour annotation sessions.
All three views update synchronously as the annotator moves through the timeline.

%-----------------------------------------------------
\section{Corpus}\label{sec:dataset}
%-----------------------------------------------------

The annotated corpus consists of all 153 piano performances of the F\"urElise dataset~\cite{wang2024furelise}, totaling $\sim$5.3 \, M note events with fingering labels.
All 153 pieces have been through a first expert review pass ($R_1$); 62 have a second pass ($R_2$) and 91 have a third pass ($R_3$).
$R_1$ passes were produced by trained research annotators; $R_2$ and $R_3$ were performed individually by a different music specialist who independently reviewed all assigned pieces.
Where the two specialists disagreed, the final label was resolved by consensus.
This two-specialist review covers 152 of 153 pieces and improves label quality, but it does not provide a formal inter-annotator agreement coefficient.
For every piece, we store \emph{paired} fingering tracks: a rule-based prediction $f_{\text{rule}}$ and a human-edited version $f_{\text{edited}}$ on the same motion-frame grid.
This pairing is the structural feature that distinguishes the corpus from prior fingering datasets: each note is labelled with both what an automatic annotator predicted and what the human ultimately accepted, so the rule-error population is a first-class object that downstream models can train on or audit against.
Aggregate statistics are in \tabref{tab:corpus}.
The 91.82\% rule--edited agreement is computed over all 5{,}301{,}309 edited notes.
This aggregate combines wrong-finger assignments with notes that the rule annotator missed entirely, both counted as rule errors.
Per-piece disagreement has median 7.4\% (IQR [4.9, 10.6]\%, max 28.4\%), leaving $\sim$433\,k rule-error notes as the natural correction target.
Unlike PIG~\cite{nakamura2020statistical}, whose labels are score-level and monolithic, PiAnnotate is frame-aligned to performance motion and stores rule and human tracks in parallel, enabling the held-out diagnostics in Sec.~\ref{sec:findings}.

\textbf{Data card.}
\emph{Source recordings} (153 pieces, $\sim$10 \, h, 15 elite pianists, Baroque through 20C) are reused from F\"urElise~\cite{wang2024furelise} and \emph{not} redistributed; users obtain them from the F\"urElise release.
\emph{What we add}: (i) the rule-based fingering track $f_{\text{rule}}$ from a geometric distance assignment on the F\"urElise hand meshes; (ii) the human-edited track $f_{\text{edited}}$ produced through the PiAnnotate tool ($R_1$ by research annotators, $R_2$/$R_3$ by two independent music specialists with consensus resolution); (iii) per-piece $R_1$/$R_2$/$R_3$ status JSONs.
The annotation tool, probe code, and analysis scripts, but not the edited fingering labels, are released under MIT at \url{https://github.com/joonhyungbae/PiAnnotate}.

\begin{table}[h]
  \centering
  \small
  \begin{tabular}{lr}
    \toprule
    Property & Value \\
    \midrule
    Pieces                          & 153 \\
    Pieces $R_1$-verified           & 153 \\
    Pieces with $R_2$ review        & 62  \\
    Pieces with $R_3$ review        & 91  \\
    Total notes (rule)              & 5{,}089{,}395 \\
    Total notes (edited)            & 5{,}301{,}309 \\
    Rule--edited agreement          & 91.82\% \\
    Rule-error population           & 8.18\% (433\,410 notes) \\
    \bottomrule
  \end{tabular}
  \caption{The PiAnnotate corpus at the time of writing. Edited contains $\sim$200\,k more notes than rule because human reviewers also \emph{add} fingerings that the rule-based annotator missed entirely.}
  \label{tab:corpus}
\end{table}

%-----------------------------------------------------
\section{Diagnostic Probe}\label{sec:probe}
%-----------------------------------------------------

We train a small causal Transformer encoder~\cite{vaswani2017attention} in pairs $(f_{\text{rule}}, f_{\text{edited}})$ from the corpus.
Let $f^{\text{rule}}_i \in \{0,1,\dots,10\}$ be the rule label for the onset of the $i$ -th note (0 = missing) and $f^\star_i$ the edited label.
Each note is encoded by a 77-d feature vector.
It contains five key-geometry terms (normalized key index, black-key flag, key-center $x/y$, and surface height), 60 fingertip-geometry terms (10 fingertips $\times$ $x/y/z$ offsets to the current key, absolute height, and two in-range flags), and 12 rule-label descriptors (hand/finger one-hots and missing/match flags).
A NoteEncoder projects these to $d$ and embeds $f^{\text{rule}}_i$ with a learnable per-class vector that is added to the note representation.
Note embeddings are pooled per onset group and fed to a causal Transformer encoder.
The main configuration used in Sec.~\ref{sec:findings} is intentionally small, with 1 encoder layer and $d{=}64$.
Its learned rule-label embedding is zeroed and frozen at training time, while fixed rule descriptors still remain in the 77-d note vector and the inference gate.
A 4-layer, $d{=}256$ configuration with the rule-label embedding active is reported as a capacity ablation.
Two heads are trained jointly: an 11-way class head and a binary correction head, with loss
\[
  \mathcal{L} = \mathrm{CE}(f^{\text{cls}}_i, f^\star_i) + \mathrm{BCE}(c_i, c^\star_i),
\]
where $c^\star_i = \mathbf{1}[f^{\text{rule}}_i \neq f^\star_i]$.
We deliberately take rule-based fingering rather than raw MIDI as input, because we observed that MIDI--motion frame alignment drifts on long pieces (the first $\sim$5 \,000 frames match at 99.6\% but later sections fall to 10--30\%); the rule-based fingering is already on the motion frame grid.

\textbf{Inference gate.}
The probe overrides the rule label for the note $i$ when
\begin{equation}\label{eq:gate}
  \big(\hat f^{\text{cls}}_i \neq f^{\text{rule}}_i\big) \;\wedge\; \big(p^{\text{cls}}_i > 0.9\big) \;\wedge\; \big(p^{\text{cls}}_i / p^{\text{rule}}_i > 2\big),
\end{equation}
where $p^{\text{cls}}_i$ and $p^{\text{rule}}_i$ are the probabilities of the classifier head for its prediction top-1 and for the rule label, respectively.
The thresholds $0.9$ and $2\times$ are implementation defaults and were not tuned on the slices reported in Sec.~\ref{sec:findings}.
A sweep of sensitivity on seed 0 (no-rule embossed) over the top-1 threshold $\tau \in \{0.70, 0.80, 0.90, 0.95\}$ gives holdout margins of $\{-2.14, +1.28, +2.78, +2.83\}$\, pp, respectively, so $0.90$ sits on the flat end of the curve and $0.95$ is nearly indistinguishable from the default.
The correction-head logit is an auxiliary training objective and does not enter \eqnref{eq:gate}; sweeping on its probability alone yields $P{\leq}0.13$.
All numbers in Sec.~\ref{sec:findings} are produced by \eqnref{eq:gate}.

%-----------------------------------------------------
\section{Findings}\label{sec:findings}
%-----------------------------------------------------

We score the probe against the current edited labels in the full corpus and in the $R_2$/$R_3$ subsets.
All 153 pieces are in training; the subsets differ in how many human passes their labels have received, so we read them as label-quality slices rather than as held-outs.

\textbf{Finding 1. Precision-first triage.}
In the whole corpus, the gate of \eqnref{eq:gate} indicates 1.91\% of the notes as suspect.
Of those flags, 79.7\% land on notes where the rule actually disagrees with the human -- about 10$\times$ above the 8.18\% chance rate.
The cost of this precision is the recall: only 28.0\% of rule errors are flagged.
The two paired numbers, in this order, are the operating point at which we recommend reading the probe.

\textbf{Finding 2. Near-zero break rate.}
In the same corpus, the probe overwrites only 14{,}011 of 4{,}867{,}897 originally-correct rule labels (0.29\%).
In the $R_2$ cut this is 6{,}876 of 2{,}297{,}010 (0.30\%).
The gate of \eqnref{eq:gate} is structurally conservative: when in doubt it leaves the rule alone.
This is the property that makes the probe usable as a triage filter rather than as a re-annotator.

\textbf{Held-out generalization.}
To separate memorization from generalizable signal, we retrained the main probe (1L, $d{=}64$, no-rule embedded) on the 91 pieces that are not in $R_2$, held out the 62 $R_2$ pieces entirely, and repeated with five independent random seeds.
We report $\Delta$ as probe accuracy minus rule accuracy on the same held-out pieces, measured in percentage points.
The per-seed margins are $\Delta \in [+2.54, +2.92]$\,pp; every paired piece-level cluster bootstrap (5 \,000 resamples of whole pieces) places its 95\% lower bound above zero.
The combination gives $\Delta = \mathbf{+2.83}$\,pp (Student $t$ CI $[+2.63, +3.03]$, $df{=}4$), with recall $42.6 \pm 2.1\%$, precision $88.4 \pm 5.4\%$, and break rate $0.19 \pm 0.33\%$.
Thus the same 62 pieces that show $-5.22$\,pp train-fit underperformance show positive held-out margins in every seed.

\textbf{Robustness.}
All four perturbations leave the conclusion intact (\tabref{tab:findings}, bottom blocks).
\emph{Capacity}: scaling to 4 layers, $d{=}256$ gives $+2.14$\,pp ($[+1.43, +2.86]$) -- the same mean, $3.6{\times}$ wider $\sigma$, so extra capacity only adds calibration noise.
\emph{Rule label as input}: reinforcing the embedding of the rule label in the larger model gives $+2.10$\,pp ($[+1.19, +3.01]$), indistinguishable in mean but with higher break rate, ruling out a rule-as-input shortcut.
\emph{Partition}: an independent random 91/62 split gives $+3.10$\,pp ($[+2.54, +3.65]$, recall $52\%$), so the $R_2$ partition is not a suspiciously easy slice.
\emph{Architecture family}: a gradient-boosted decision tree (GBDT)~\cite{friedman2001greedy}, implemented as a per-note classifier on the same 77-d features without sequence context, reaches only $70.4\%$ on the $R_2$ holdout under \eqnref{eq:gate}; even with $\tau{=}0.99$, its margin is $\Delta=-2.94$\,pp.
The signal therefore requires the per-note features plus minimal causal attention.

\textbf{Label vintage.}
The train-fit agreement is $+1.18$\,pp on $R_3$ but $-5.22$\,pp on $R_2$ (\tabref{tab:findings}).
A timestamp audit explains this gap.
For 50/59 $R_2$ pieces, the probe output predates the $R_2$ completion timestamp, i.e.\ the labels against which it is scored did not yet exist when it was trained.
The lesson is operational: paired-track corpora under active editing need explicit label-vintage tracking, which we now record in the per-piece status JSON.

\begin{table}[t]
  \centering
  \footnotesize
  \begin{tabular}{lcccc}
    \toprule
    Setting & Flag P/R & Break & Rule & Probe \\
    \midrule
    \multicolumn{5}{l}{\emph{Train-fit, single seed (all 153 in training)}} \\
    All $R_1$ (153) & 79.7/28.0 & 0.29\% & 91.82 & 89.98 \\
    $R_3$ (91)      & 81.8/33.3 & 0.28\% & 92.39 & \textbf{93.57} \\
    $R_2$ (62)      & 76.9/22.9 & 0.30\% & 91.39 & 86.17 \\
    \midrule
    \multicolumn{5}{l}{\emph{Held-out 5-seed (62 $R_2$ pieces excluded)}} \\
    Main (1L,64)   & 88.4/42.6 & 0.19\% & 91.39 & \textbf{94.22} \\
    Abl.\ 4L,256   & 81.5/43.1 & 0.58\% & 91.39 & 93.54 \\
    Abl.\ +rule    & 78.6/43.1 & 0.85\% & 91.39 & 93.49 \\
    \midrule
    \multicolumn{5}{l}{\emph{Held-out 5-seed (random 62/91 split, 4L)}} \\
    Random         & 85.8/52.2 & 0.47\% & 91.28 & \textbf{94.37} \\
    \midrule
    \multicolumn{5}{l}{\emph{Non-sequence baseline (GBDT, 1 seed, $R_2$ split)}} \\
    GBDT 77-d feats & 12.9/79.8 & 26.5\% & 91.39 & 70.41 \\
    \bottomrule
  \end{tabular}
  \caption{Probe diagnostics. The top block is train-fit on label-quality slices, not a generalisation test. The middle block is the 5-seed $R_2$-excluded holdout; the main probe freezes the learned rule-label embedding, while fixed rule descriptors and the gate remain. Bottom blocks report a random split and GBDT baseline.}
  \label{tab:findings}
\end{table}

%-----------------------------------------------------
\section{Discussion and Conclusion}\label{sec:discussion}
%-----------------------------------------------------

PiAnnotate does not remove expert review, and the probe is not a deployable corrector.
Its role is diagnostic: paired rule/edited annotations support a conservative triage point, a near-zero break rate, and a $+2.83$\,pp held-out gain.
The capacity, random-split, and GBDT checks argue against a simple learned rule-label shortcut.

\textbf{Limitations.}
The corpus uses one capture setup and 15 performers; no formal inter-annotator agreement coefficient is reported.
The held-out split follows review stage rather than composer or difficulty, reviewer-time impact is unmeasured, and the no-rule-embedding ablation still leaves fixed rule descriptors and the gate.
The released artifacts are the tool and probe code, not the edited labels.
Together, these results position PiAnnotate as an inspectable expert-annotation workflow; future work should measure reviewer time and evaluate beyond F\"urElise.

% Bibliography
\clearpage
\bibliography{KSMI_template}

@inproceedings{wang2024furelise,
  title={F{\"u}relise: Capturing and physically synthesizing hand motion of piano performance},
  author={Wang, Ruocheng and Xu, Pei and Shi, Haochen and Schumann, Elizabeth and Liu, C Karen},
  booktitle={SIGGRAPH Asia 2024 Conference Papers},
  pages={1--11},
  year={2024}
}

@article{nakamura2020statistical,
  title={Statistical learning and estimation of piano fingering},
  author={Nakamura, Eita and Saito, Yasuyuki and Yoshii, Kazuyoshi},
  journal={Information Sciences},
  volume={517},
  pages={68--85},
  year={2020},
  publisher={Elsevier}
}

@article{parncutt1997model,
  title={An ergonomic model of keyboard fingering for melodic fragments},
  author={Parncutt, Richard and Sloboda, John A and Clarke, Eric F and Raekallio, Matti and Desain, Peter},
  journal={Music Perception},
  volume={14},
  number={4},
  pages={341--382},
  year={1997},
  publisher={University of California Press}
}

@inproceedings{ramoneda2022automatic,
  title={Automatic piano fingering from partially annotated scores using autoregressive neural networks},
  author={Ramoneda, Pedro and Jeong, Dasaem and Nakamura, Eita and Serra, Xavier and Miron, Marius},
  booktitle={Proceedings of the 30th ACM International Conference on Multimedia},
  pages={6502--6510},
  year={2022}
}

@article{romero2017embodied,
  title={Embodied hands: Modeling and capturing hands and bodies},
  author={Romero, Javier and Tzionas, Dimitrios and Black, Michael J},
  journal={ACM Transactions on Graphics},
  volume={36},
  number={6},
  pages={245:1--245:17},
  year={2017}
}

@article{vaswani2017attention,
  title={Attention is all you need},
  author={Vaswani, Ashish and Shazeer, Noam and Parmar, Niki and Uszkoreit, Jakob and Jones, Llion and Gomez, Aidan N and Kaiser, {\L}ukasz and Polosukhin, Illia},
  journal={Advances in neural information processing systems},
  volume={30},
  year={2017}
}

@techreport{settles2009active,
  title={Active learning literature survey},
  author={Settles, Burr},
  year={2009},
  institution={University of Wisconsin-Madison},
  number={Computer Sciences Technical Report 1648}
}

@inproceedings{kim2025pianovam,
  author    = {Kim, Yonghyun and Park, Junhyung and Bae, Joonhyung
               and Kim, Kirak and Kwon, Taegyun and Lerch, Alexander
               and Nam, Juhan},
  title     = {{PianoVAM}: A Multimodal Piano Performance Dataset},
  booktitle = {Proc. Int. Soc. Music Information Retrieval Conf. (ISMIR)},
  year      = {2025}
}

@misc{park2025webtoolkits,
  author        = {Park, Junhyung and Kim, Yonghyun and Bae, Joonhyung
                   and Kim, Kirak and Kwon, Taegyun and Lerch, Alexander
                   and Nam, Juhan},
  title         = {Two Web Toolkits for Multimodal Piano Performance Dataset
                   Acquisition and Fingering Annotation},
  year          = {2025},
  eprint        = {2509.15222},
  archivePrefix = {arXiv},
  primaryClass  = {cs.SD}
}

@article{friedman2001greedy,
  author  = {Friedman, Jerome H.},
  title   = {Greedy Function Approximation: A Gradient Boosting Machine},
  journal = {The Annals of Statistics},
  volume  = {29},
  number  = {5},
  pages   = {1189--1232},
  year    = {2001}
}

@article{northcutt2021confident,
  title={Confident learning: Estimating uncertainty in dataset labels},
  author={Northcutt, Curtis and Jiang, Lu and Chuang, Isaac},
  journal={Journal of Artificial Intelligence Research},
  volume={70},
  pages={1373--1411},
  year={2021}
}

@inproceedings{cannam2010sonic,
  title={Sonic visualiser: An open source application for viewing, analysing, and annotating music audio files},
  author={Cannam, Chris and Landone, Christian and Sandler, Mark},
  booktitle={Proceedings of the 18th ACM international conference on Multimedia},
  pages={1467--1468},
  year={2010}
}

@inproceedings{brugman2004elan,
  title={Annotating multimedia/multi-modal resources with ELAN},
  author={Brugman, Hennie and Russel, Albert},
  booktitle={Proceedings of the 4th International Conference on Language Resources and Evaluation (LREC)},
  year={2004}
}

@inproceedings{alkasimi2007physical,
  author    = {Al Kasimi, Amina and Nichols, Eric and Raphael, Christopher},
  title     = {A Simple Algorithm for Automatic Generation of Polyphonic Piano Fingerings},
  booktitle = {Proc. 8th Int. Society for Music Information Retrieval Conf. (ISMIR)},
  pages     = {355--356},
  year      = {2007},
  address   = {Vienna, Austria}
}

@article{belinkov2019probing,
  title={Analysis methods in neural language processing: A survey},
  author={Belinkov, Yonatan and Glass, James},
  journal={Transactions of the Association for Computational Linguistics},
  volume={7},
  pages={49--72},
  year={2019}
}

@article{song2022learning,
  title={Learning from noisy labels with deep neural networks: A survey},
  author={Song, Hwanjun and Kim, Minseok and Park, Dongmin and Shin, Yooju and Lee, Jae-Gil},
  journal={IEEE transactions on neural networks and learning systems},
  volume={34},
  number={11},
  pages={8135--8153},
  year={2022},
  publisher={IEEE}
}

@inproceedings{yonebayashi2007automatic,
  title={Automatic Decision of Piano Fingering Based on a Hidden Markov Models.},
  author={Yonebayashi, Yuichiro and Kameoka, Hirokazu and Sagayama, Shigeki},
  booktitle={IJCAI},
  volume={7},
  pages={2915--2921},
  year={2007}
}

@inproceedings{huang2019music,
  title={Music transformer},
  author={Huang, Cheng-Zhi Anna and Vaswani, Ashish and Uszkoreit, Jakob and Shazeer, Noam and Simon, Ian and Hawthorne, Curtis and Dai, Andrew M and Hoffman, Matthew D and Dinculescu, Monica and Eck, Douglas},
  booktitle={International Conference on Learning Representations},
  year={2019}
}

@inproceedings{simon2017hand,
  title={Hand keypoint detection in single images using multiview bootstrapping},
  author={Simon, Tomas and Joo, Hanbyul and Matthews, Iain and Sheikh, Yaser},
  booktitle={Proceedings of the IEEE conference on Computer Vision and Pattern Recognition},
  pages={1145--1153},
  year={2017}
}

@article{goebl2013temporal,
  title={Temporal control and hand movement efficiency in skilled music performance},
  author={Goebl, Werner and Palmer, Caroline},
  journal={PloS one},
  volume={8},
  number={1},
  pages={e50901},
  year={2013},
  publisher={Public Library of Science San Francisco, USA}
}

@article{boersma2001praat,
  title={Praat, a system for doing phonetics by computer},
  author={Boersma, Paul},
  journal={Glot. Int.},
  volume={5},
  number={9},
  pages={341--345},
  year={2001}
}

\end{document}